\documentclass[12pt]{article}
\usepackage{amsmath,amssymb,amsthm,amsxtra,overpic,bbm,bm,epsfig,subfigure}
\usepackage{hyperref}
\usepackage{graphicx}
\usepackage{color}
\usepackage{comment}
\usepackage{epstopdf}
\usepackage{float}
\usepackage{cite}
\usepackage{overpic} 
\usepackage{multirow}
\usepackage{pifont}

\usepackage[title]{appendix}

\usepackage{hyperref}
\usepackage{url}
\usepackage{slashed,stmaryrd}

\usepackage{lscape}
\usepackage{array}
\usepackage{booktabs}
\newcommand{\tabincell}[2]{\begin{tabular}{@{}#1@{}}#2\end{tabular}}

\def\thefootnote{\fnsymbol{footnote}}

\def\thefootnote{\fnsymbol{footnote}}

\begin{document}
	
\vspace{0.2cm}
	
\begin{center}
{\large\bf A modular $A_4$ symmetry realization of two-zero textures of the Majorana neutrino mass matrix}
\end{center}
	
\vspace{0.2cm}
	
\begin{center}
{\small \bf Di Zhang$^{a,b}$}\footnote{Email: zhangdi@ihep.ac.cn} \\
{$^a$Institute of High Energy Physics, Chinese Academy of Sciences, Beijing 100049, China \\
$^b$School of Physical Sciences, University of Chinese Academy of Sciences, Beijing 100049, China}
\end{center}

\vspace{1.5cm}
	
\begin{abstract}
We show how to realize two-zero textures of the Majorana neutrino mass matrix $M_\nu$ based on modular $A_4$ invariant models without flavons. In these models, all matter fields are assigned to three inequivalent singlets, ${\bf 1}$, ${\bf 1^\prime}$ and ${\bf 1^{\prime\prime}}$, of the finite modular group $\Gamma_3 \simeq A_4$. Considering tensor products of the $A_4$ group, it is easy to make the charged lepton mass matrix $M_\ell$ diagonal. Since not all modular forms of a specific weight and level 3 can be arranged into three inequivalent singlets of $A_4$ simultaneously, we can always make some entries in $M_\nu$ vanish by properly assigning the representations and modular weights for the matter fields. We consider two cases where neutrino masses originate from the Weinberg operator and the type-\uppercase\expandafter{\romannumeral1} seesaw mechanism, respectively. For the former case, all seven viable two-zero textures of $M_\nu$ (${\bf A_{1,2}}$, ${\bf B_{1,2,3,4}}$ and ${\bf C}$) can be realized successfully. For the latter case, only five of them (namely ${\bf A_{1,2}}$, ${\bf B_{3,4}}$ and ${\bf C}$) can be achieved due to the intrinsic structure of the right-handed neutrino mass matrix $M_{\rm R}$ in our assumption for simplicity. 
\end{abstract}
	
	
\newpage

\def\thefootnote{\arabic{footnote}}
\setcounter{footnote}{0}
\setcounter{table}{0}
\setcounter{equation}{0}
\setcounter{figure}{0}

\section{Introduction}

Compelling evidences obtained from solar, atmospheric, reactor and accelerator neutrino experiments in the last two decades~\cite{Tanabashi:2018oca} have proved the existence of neutrino oscillations~\cite{Pontecorvo:1957cp,Pontecorvo:1957qd,Maki:1962mu}, implying that neutrinos have nonzero and nondegenerate masses, and flavor mixing in the lepton sector exists. The standard model (SM) itself tells us nothing about the quantitative details of Yukawa interactions, thus it is rather challenging to explore the underlying flavor structures of charged fermions and massive neutrinos. It calls for new physics beyond the SM to control the Yukawa couplings of quarks and leptons and understand flavor mixing. 
In view of the fact that a convincing flavor theory is lacking, approaches such as flavor symmetries and texture zeros or their combinations have been widely explored to shed light on the flavor secrets of fermions.

The flavor symmetry with a non-Abelian discrete group is a popular approach to explain lepton flavor mixing pattern (see reviews~\cite{Altarelli:2010gt,Ishimori:2010au,King:2013eh,Petcov:2017ggy} and references therein). 
Recently, a new and attractive approach with the modular symmetry applied to the lepton flavor problem has been proposed in~\cite{Feruglio:2017spp}. In such a modular invariant model, only a few flavons or even no flavons need to be introduced and the Yukawa couplings or the right-handed neutrino mass matrix in the type-\uppercase\expandafter{\romannumeral1} seesaw mechanism~\cite{Minkowski:1977sc,Yanagida:1979as,GellMann:1980vs,Glashow:1979nm,Mohapatra:1979ia} are regarded as modular forms which are holomorphic functions of the modulus $\tau$  and transform non-trivially under the modular symmetry. In the model without flavons, the vacuum expectation value (VEV) of the modulus $\tau$ is the only source of symmetry breaking. In the light of these advantages, a lot of works have been done and successfully predicted neutrino masses and mixing parameters in terms of a few input parameters, such as the models with the modular $\Gamma_2 \simeq S_3$~\cite{Kobayashi:2018vbk,Okada:2019xqk}, $\Gamma_3 \simeq A_4$~\cite{Kobayashi:2018vbk,Criado:2018thu,Kobayashi:2018scp,Novichkov:2018yse,Nomura:2019jxj,Nomura:2019yft,Ding:2019zxk,Okada:2019mjf,Nomura:2019lnr,Asaka:2019vev,Gui-JunDing:2019wap}, $\Gamma_4 \simeq S_4$~\cite{Gui-JunDing:2019wap,Penedo:2018nmg,Novichkov:2018ovf,Novichkov:2019sqv,deMedeirosVarzielas:2019cyj,Kobayashi:2019mna,King:2019vhv,Okada:2019lzv,Criado:2019tzk,Kobayashi:2019xvz} or $\Gamma_5 \simeq A_5$~\cite{Novichkov:2018nkm,Ding:2019xna,Criado:2019tzk} symmetry. This approach has also been extended to the quark sector~\cite{deAnda:2018ecu,Okada:2018yrn,Kobayashi:2018wkl,Okada:2019uoy,Kobayashi:2019rzp}.

Assuming that neutrinos are Majorana particles and we work in the flavor basis where the charged lepton mass matrix $M_\ell$ is diagonal, the textures of the Majorana neutrino mass matrix $M_\nu$ with more than two zeros~\footnote{Owing to the symmetric structure of the Majorana neutrino mass matrix, the vanishing off-diagonal elements (m,n) and (n,m) are taken to be one texture zero.} are not compatible with current experimental results~\cite{Xing:2004ik} and the predictive power of one-zero textures of $M_\nu$ is limited~\cite{Xing:2003jf,Xing:2003ic,Merle:2006du,BenTov:2011tj,Lashin:2011dn,Gautam:2015kya,Gautam:2018izb,Singh:2018tqu,Kitabayashi:2018bye,Liu:2018cka,Kitabayashi:2018jnl}. In contrast, two-zero textures of $M_\nu$ put forward in 2002~\cite{Frampton:2002yf,Xing:2002ta,Xing:2002ap} are more fascinating since in these cases, the lightest neutrino mass ($m_1$ for the normal mass order or $m_3$ for the inverted mass order) and two Majorana phases $\rho$ and $\sigma$ can be determined by six neutrino oscillation parameters (i.e., $\Delta m^2_{21}$, $\Delta m^2_{31}$, $\theta_{12}$, $\theta_{13}$, $\theta_{23}$ and the Dirac phase $\delta_{\rm CP}$)~\cite{Xing:2002ta,Xing:2002ap,Guo:2002ei,Xing:2003ez,Xing:2019vks}. There are totally fifteen two-zero textures of the Majorana neutrino mass matrix $M_\nu$, while only seven of them, namely ${\bf A_{1,2}}$, ${\bf B_{1,2,3,4}}$ and ${\bf C}$, are allowed by current experimental data within $3\sigma$. The survival statues and phenomenological implications of these two-zero textures have been extensively studied in~\cite{Frampton:2002rn,Kageyama:2002zw,Desai:2002sz,Frigerio:2002fb,Kaneko:2002yp,Bhattacharyya:2002aq,Honda:2003pg,Dev:2006qe,Branco:2007nn,Grimus:2011sf,Fritzsch:2011qv,Meloni:2014yea,Cebola:2015dwa,Zhou:2015qua,Singh:2016qcf,Alcaide:2018vni,Singh:2019baq,Borgohain:2019pya}. Moreover, it is well known that zero textures of $M_\nu$ can be always realized with Abelian flavor symmetries~\cite{Grimus:2004hf}. The realizations of viable two-zero textures with Abelian flavor symmetries are also discussed in~\cite{Fritzsch:2011qv,Grimus:2004az,Dev:2011jc,Felipe:2014vka,Borgohain:2018lro}. In ref.~\cite{Hirsch:2007kh}, a predictive model with the non-Abelian $A_4$ flavor symmetry is presented, which can achieve the textures ${\bf B_{1,2}}$. And the models with the $A_4$ flavor symmetry giving the textures ${\bf A_{1,2}}$ and ${\bf B_4}$ are proposed in~\cite{Zhou:2015qua}. Similar realizations of two-zero textures with the $A_4$ group are also presented in~\cite{Lamprea:2016egz,delaVega:2018cnx}. However, in these flavor symmetry models, a generous set of scalar multiplets, the so-called flavons, have to be introduced and the associated potential structures have to be constructed carefully.

In the present paper, we impose the modular $A_4$ symmetry on the action of the lepton sector in an $\mathcal{N}=1$ supersymmetry model, where no flavons need to be introduced and all chiral superfields are assigned to three inequivalent singlets of the $A_4$ group, i.e., ${\bf 1}$, ${\bf 1^\prime}$ and ${\bf 1^{\prime\prime}}$. With the help of tensor products of the $A_4$ group, it is easy to obtain the diagonal charged lepton mass matrix. Since not all modular forms of a specific weight and level 3 can be arranged into three inequivalent singlets of $A_4$ simultaneously, there are enough degrees of freedom to properly assign the representations and modular weights for the chiral superfields, leading to two-zero textures of $M_\nu$. It is worth pointing out that arranging only three inequivalent singlets of the $A_4$ group is equivalent to considering a $Z_3$ symmetry and to some extent, the application of the modular weights is phenomenologically similar to imposing an additional Abelian symmetry. In this work, we consider neutrino masses that originate either from the Weinberg operator~\cite{Weinberg:1979sa} or from the type-\uppercase\expandafter{\romannumeral1} seesaw mechanism~\cite{Minkowski:1977sc,Yanagida:1979as,GellMann:1980vs,Glashow:1979nm,Mohapatra:1979ia}. In the former case, all seven allowed two-zero textures of $M_\nu$, ${\bf A_{1,2}}$, ${\bf B_{1,2,3,4}}$ and ${\bf C}$, can be realized successfully. In the latter case, only five of them, ${\bf A_{1,2}}$, ${\bf B_{3,4}}$ and ${\bf C}$, can be achieved, as constrained by the intrinsic structure of the right-handed neutrino mass matrix $M_{\rm R}$ under discussion. 

The rest of this paper is organized as follows. In section 2, we briefly introduce the supersymmetry model with the modular $A_4$ symmetry and derive modular forms of weights up to 10 and level 3. Then we show the realizations of two-zero textures with modular $A_4$ invariant models where neutrino masses originate either from the Weinberg operator or from the type-\uppercase\expandafter{\romannumeral1} seesaw mechanism in section 3. The summary and remarks are given in section 4.

\def\thefootnote{\arabic{footnote}}
\setcounter{footnote}{0}
\setcounter{table}{0}
\setcounter{equation}{0}
\setcounter{figure}{0}

\section{Modular $A_4$ Symmetry and Modular Forms}
The infinite modular group $\overline{\Gamma}$ is defined by:
\begin{eqnarray}
\overline{\Gamma} = \left\{ \left(\begin{matrix}
a & b \\ c & d
\end{matrix}\right)/(\pm 1), \quad a,b,c,d \in \mathbb{Z},\quad ad - bc = 1 \right\} \;,
\label{MG:def}
\end{eqnarray}
where $\mathbb{Z}$ stands for the set of all integers and $\overline{\Gamma}$ acts on the complex modulus $\tau$ with ${\rm Im}{\tau} > 0$ as linear fractional transformations: 
\begin{eqnarray}
\gamma:~ \tau \rightarrow \gamma\tau = \frac{a\tau + b}{c\tau + d} \;,
\label{MG:tran}
\end{eqnarray}
in which $\gamma$ is an arbitrary element of the $\overline{\Gamma}$ group. The modular group has two generators, denoted as $S$ and $T$, satisfying $S^2 = (ST)^3 = \mathbbm{1}$. The transformations $S$ and $T$ can be expressed as
\begin{eqnarray}
S = \left(\begin{matrix} 0 & 1 \\ -1 & 0 \end{matrix}\right) \;,\quad T = \left(\begin{matrix} 1 & 1 \\ 0 & 1 \end{matrix}\right)
\label{MG:rep}
\end{eqnarray}
and then the modulus $\tau$ transforms under $S$ and $T$ as
\begin{eqnarray}
S:~ \tau \rightarrow - \frac{1}{\tau} \;,\quad T:~ \tau \rightarrow \tau + 1 \;,
\label{MG:gtran}
\end{eqnarray}
respectively. The modular group $\overline{\Gamma}$ is isomorphic to the projective special linear group $PSL(2,\mathbb{Z})$, which is the quotient group of the 2-dimensional special linear group $SL(2,\mathbb{Z})$ by its center $\mathbb{Z}_2 = \{ \mathbbm{1},-\mathbbm{1}\}$. A series of normal subgroups of $SL(2,\mathbb{Z})$ can be denoted as
\begin{eqnarray}
\Gamma(N) = \left\{\left(\begin{matrix} a & b \\ c & d \end{matrix}\right) \in SL(2,\mathbb{Z}),\quad \left(\begin{matrix} a & b \\ c & d \end{matrix}\right) = \left(\begin{matrix} 1 & 0 \\ 0 & 1 \end{matrix}\right) \left( {\rm mod} ~ N\right) \right\}
\label{MGN}
\end{eqnarray}
with $N \in \{x \geq 2 ~|~ x \in \mathbb{Z}\}$ and these normal subgroups are all infinite groups. Let us define $\overline{\Gamma}(2) \equiv \Gamma(2) / \mathbb{Z}_2$ for $N=2$ and $\overline{\Gamma}(N) \equiv \Gamma(N)$ for $N>2$, since $\mathbb{Z}_2$ is not the subgroup of $\Gamma(N)$ with $N>2$. The quotient groups of the modular group $\overline{\Gamma}$ by $\overline{\Gamma}(N)$, labeled as $\Gamma_N \equiv \overline{\Gamma} / \overline{\Gamma}(N)$, are called finite modular group, in which $T^{N} = \mathbbm{1}$ is satisfied except for the relation $S^2 = (ST)^3 = \mathbbm{1}$. For $N=2,3,4,5$ the isomorphisms, $\Gamma_2 \simeq S_3$, $\Gamma_3 \simeq A_4$, $\Gamma_4 \simeq S_4$, $\Gamma_5 \simeq A_5$, hold respectively.

Modular forms $Y(\tau)$ of non-negative and even weight $k$~\footnote{In ref.~\cite{Liu:2019khw}, this modular invariant approach has been extended to include odd weight modular forms.} and level $N$ are holomorphic functions of the modulus $\tau$, and they span a linear space whose dimension is dependent on its weight $k$ and level $N$~\cite{Feruglio:2017spp}. In this linear space, it is always possible to choose a basis where modular forms transform under the $\Gamma_N$ group as
\begin{eqnarray}
Y_i(\gamma \tau) = (c\tau + d)^k \rho(\gamma)_{ij} Y_j(\tau) \;,
\label{MF:tran}
\end{eqnarray}
where $\gamma \in \Gamma_N$ and $\rho$ is the unitary representation of $\Gamma_N$.

Modular forms play a very significant role in model building to explain fermion masses and flavor mixing~\cite{Feruglio:2017spp}. For the case of level 3 corresponding to the group $\Gamma_3 \simeq A_4$, the dimension of the linear space spanned by modular forms of weight $k$ is $k+1$. Therefore, for $k=0$, there is only one trivial modular form which is a constant and independent of the modulus $\tau$. For $k=2$, there are three linearly independent modular forms forming a triplet of $A_4$. Then the modular forms of higher weights can be constructed by the products of those of weight 2. To obtain the right number of linearly independent modular forms of higher weights, viz., the right dimension of linear space, there are usually some relations between the products of the 
modular forms of the lowest non-trivial weight~\cite{Feruglio:2017spp}.

The modular forms of weight 2 and level 3 can be constructed by the Dedekind eta-function which is defined in the upper complex plane as
\begin{eqnarray}
\eta(\tau) = q^{1/24} \prod^{\infty}_{n=1} (1- q^n)
\label{MF:eta}
\end{eqnarray}
with $q = e^{i2\pi\tau}$, and written in term of the Dedekind eta-function and its derivative as~\cite{Feruglio:2017spp}
\begin{eqnarray}
Y^{}_{1}(\tau) &=& \frac{i}{2\pi} \left[ \frac{\eta^{\prime}(\tau/3)}{\eta(\tau/3)} + \frac{\eta^{\prime}((\tau+1)/3)}{\eta((\tau+1)/3)} +\frac{\eta^{\prime}((\tau+2)/3)}{\eta((\tau+2)/3)} -\frac{27\eta^{\prime}(3\tau)}{\eta(3\tau)} \right] \;,
\nonumber
\\
Y^{}_{2}(\tau) &=& -\frac{i}{\pi} \left[ \frac{\eta^{\prime}(\tau/3)}{\eta(\tau/3)} + \omega^2 \frac{\eta^{\prime}((\tau+1)/3)}{\eta((\tau+1)/3)} + \omega \frac{\eta^{\prime}((\tau+2)/3)}{\eta((\tau+2)/3)} \right] \;,
\nonumber
\\
Y^{}_{3}(\tau) &=& -\frac{i}{\pi} \left[ \frac{\eta^{\prime}(\tau/3)}{\eta(\tau/3)} + \omega \frac{\eta^{\prime}((\tau+1)/3)}{\eta((\tau+1)/3)} + \omega^2 \frac{\eta^{\prime}((\tau+2)/3)}{\eta((\tau+2)/3)} \right] \;,
\label{MF:form}
\end{eqnarray}
where $\omega=e^{i2\pi/3}$. They satisfy the constraint~\cite{Feruglio:2017spp}
\begin{eqnarray}
Y^{2}_{2} + 2 Y^{}_{1} Y^{}_{3} = 0 \;.
\label{MF:cons}
\end{eqnarray}
These three modular forms can be organized into a triplet of $A_4$ transforming in the ${\bf 3 }$ irreducible representation of $A_4$, denoted as
\begin{eqnarray}
Y^{(2)}_{\bf 3} = \left(\begin{matrix} Y_1 \\ Y_2 \\ Y_3 \end{matrix}\right)\;,
\label{MF:2}
\end{eqnarray}
in which the superscript $(2)$ and subscript ${\bf 3}$ characterize the corresponding weight and multiplet of $A_4$. With the help of Eqs.~(\ref{MF:cons}) and (\ref{MF:2}), the modular forms of higher weights can be constructed by means of tensor products of $A_4$ which are given in Appendix A. Here we only give modular forms of weight up to 10 which are relevant to the following models. For $k=4$, there are 5 linearly independent modular forms, arranged into two singlets ${\bf 1}$, ${\bf 1^\prime}$ and one triplet ${\bf 3}$ of $A_4$:
\begin{eqnarray}
Y^{(4)}_{\bf 1} = Y^2_1 + 2Y_2Y_3 \;, \quad Y^{(4)}_{\bf 1^\prime} = Y^2_3 + 2Y_1Y_2 \;, \quad Y^{(4)}_{\bf 3} = \left(\begin{matrix} Y^2_1 - Y_2Y_3 \\ Y^2_3 - Y_1Y_2 \\ Y^2_2 - Y_1Y_3  \end{matrix} \right) \;.
\label{MF:4}
\end{eqnarray}
For $k=6$, one has 
\begin{eqnarray}
Y^{(6)}_{\bf 1} &=& Y^3_1 + Y^3_2 + Y^3_3 -3Y_1Y_2Y_3 \;,
\nonumber
\\
Y^{(6)}_{{\bf 3},1} &=& (Y^2_1 + 2Y_2Y_3) \left(\begin{matrix} Y_1 \\ Y_2 \\ Y_3 \end{matrix}\right) \;,\quad Y^{(6)}_{{\bf 3},2} = (Y^2_3 + 2Y_1Y_2) \left(\begin{matrix} Y_3 \\ Y_1 \\ Y_2 \end{matrix}\right)
\label{MF:6}
\end{eqnarray}
and the total dimension is 7. For $k=8$, there are
\begin{eqnarray}
Y^{(8)}_{\bf 1} &=& (Y^2_1 + 2Y_2Y_3)^2 \;,\quad Y^{(8)}_{\bf 1^\prime} = (Y^2_1 + 2Y_2Y_3) (Y^2_3 + 2Y_1Y_2) \;,\quad Y^{(8)}_{\bf 1^{\prime\prime}} = (Y^2_3 + 2Y_1Y_2)^2 \;,
\nonumber
\\
Y^{(8)}_{{\bf 3},1} &=& (Y^2_1 + 2Y_2Y_3) \left(\begin{matrix} Y^2_1 - Y_2Y_3 \\ Y^2_3 - Y_1Y_2 \\ Y^2_2 - Y_1Y_3 \end{matrix}\right) \;,\quad Y^{(8)}_{{\bf 3},2} = (Y^2_3 + 2Y_1Y_2) \left(\begin{matrix} Y^2_2 - Y_1Y_3 \\ Y^2_1 -Y_2Y_3 \\ Y^2_3 - Y_1Y_2 \end{matrix}\right) \;, 
\label{MF:8}
\end{eqnarray}
corresponding to a total dimension of 9. For $k=10$, a total of 11 linearly independent modular forms are arranged into $A_4$ multiplets as 
\begin{eqnarray}
Y^{(10)}_{\bf 1} &=& (Y^2_1 + 2Y_2Y_3) (Y^3_1 + Y^3_2 + Y^3_3 - 3Y_1Y_2Y_3) \;,
\nonumber
\\
Y^{(10)}_{\bf 1^\prime} &=& (Y^2_3 + 2Y_1Y_2) (Y^3_1 + Y^3_2 + Y^3_3 - 3Y_1Y_2Y_3) \;,
\nonumber
\\
Y^{(10)}_{{\bf 3},1} &=& (Y^2_1 + 2Y_2Y_3)^2 \left(\begin{matrix} Y_1 \\ Y_2 \\ Y_3 \end{matrix}\right) ,\quad Y^{(10)}_{{\bf 3},2} =  (Y^2_3 + 2Y_1Y_2)^2 \left(\begin{matrix} Y_2 \\ Y_3 \\ Y_1 \end{matrix}\right) \;,
\nonumber
\\
Y^{(10)}_{{\bf 3},3} &=& (Y^2_1 + 2Y_2Y_3) (Y^2_3 + 2Y_1Y_2) \left(\begin{matrix} Y_3 \\ Y_1 \\ Y_2 \end{matrix}\right) \;. 
\label{MF:10}
\end{eqnarray}
Modular forms of weights 2, 4, 6 and level 3 in Eqs.~(\ref{MF:2})---(\ref{MF:6}) have been given in~\cite{Feruglio:2017spp} while those of weights 8, 10 and level 3 in Eqs.~(\ref{MF:8})---(\ref{MF:10}) are given here for the first time. We see that only at weight $k=8$, modular forms can give all three independent singlets ${\bf 1}$, ${\bf 1^\prime}$, $\bf 1^{\prime\prime}$, while at other weights up to 10, these three singlets can not be given at the same time or there is even no singlet formed by modular forms of weight $k=2$. As one can see in section 3, this fact is quite important for the realizations of two-zero textures of the Majorana neutrino mass matrix.

Considering an ${\mathcal N} = 1$ global supersymmetry model with the modular symmetry, the general form of matter action is~\cite{Ferrara:1989bc,Ferrara:1989qb}
\begin{eqnarray}
{\mathcal S} = \int {\rm d}^4x {\rm d}^2\theta {\rm d}^2\overline{\theta} K( \phi_i, \overline{\phi}_i; \tau, \overline{\tau}) + \int {\rm d}^4x {\rm d}^2\theta W(\phi_i; \tau) + {\rm h.c.} \;,
\label{Action}
\end{eqnarray}
where $\phi_i$ is the matter chiral superfield. Under the finite modular group $\Gamma_N$, the modulus $\tau$ transforms as shown in Eq.~(\ref{MG:tran}) and the chiral superfield $\phi_i$ transforms as~\cite{Ferrara:1989bc}
\begin{eqnarray}
\gamma:~ \phi^{}_i \rightarrow (c\tau + d)^{-k_i}_{} \rho^{}_{I_i}(\gamma) \phi^{}_i \;,
\label{MG:csft}
\end{eqnarray}
where $I_i$ and $-k_i$ stand for the representation and modular weight of  the chiral superfield $\phi_i$ respectively. Since the chiral superfield $\phi_i$ is not a modular form, its weight $-k_i$ is not limited to even non-negative integer. To keep the action ${\mathcal S}$ invariant under the transformations given in Eqs.~(\ref{MG:tran}) and (\ref{MG:csft}), the superpotential $W(\phi_i;\tau)$ and the K$\ddot{\rm a}$hler potential should transform as~\cite{Ferrara:1989bc}
\begin{eqnarray}
&W(\phi_i;\tau) \rightarrow W(\phi_i;\tau) \;,&
\nonumber
\\
&K(\phi_i,\overline{\phi}_i;\tau,\overline{\tau}) \rightarrow K(\phi_i,\overline{\phi}_i;\tau,\overline{\tau}) + f(\phi_i; \tau) + \overline{f}( \overline{\phi}_i; \overline{\tau}) \;.&
\label{Pot}
\end{eqnarray}
As an example, the K$\ddot{\rm a}$hler potential in the following form~\cite{Feruglio:2017spp}~\footnote{Recently, in ref.~\cite{Chen:2019ewa}, the authors point out that modular symmetries do not fix the form of the K$\ddot{\rm a}$hler potential and the form in Eq.~(\ref{Kp}) is not complete. The full K$\ddot{\rm a}$hler potential contains additional terms which will introduce additional parameters and reduce the predictive power of these modular invariant models. For simplicity, we only consider the minimal form of the K$\ddot{\rm a}$hler potential given in Eq.~(\ref{Kp}) in this work.},
\begin{eqnarray}
K(\phi_i,\overline{\phi}_i;\tau,\overline{\tau}) = -h \ln(-i\tau +i\overline{\tau}) + \sum_{i} (-i\tau + i\overline{\tau})^{-k_i} |\phi_i|^2
\label{Kp}
\end{eqnarray}
with $h$ being a positive constant, can satisfy Eq.~(\ref{Pot}) under the transformations given by Eqs.~(\ref{MG:tran}) and (\ref{MG:csft}). After the modulus $\tau$ gains a VEV, the K$\ddot{\rm a}$hler potential in Eq.~(\ref{Kp}) gives kinetic terms for scalar components of the modulus $\tau$ and the supermultiplets $\phi_i$.

The superpotential $W(\phi_i;\tau)$ can be written in powers of the supermultiplets $\phi_i$ as
\begin{eqnarray}
W(\phi_i,\tau) = \sum_{n} \sum_{\{i_1,\cdots,i_n\}} \sum_{I_Y} (Y^{(k_Y)}_{I_Y} \phi_{i_1} \cdots \phi_{i_n})^{}_{\bf 1} \;,
\label{Sp}
\end{eqnarray}
where ${\bf 1}$ denotes the invariant singlet of the modular group $\Gamma_N$. If the superpotential is required to transform as shown in Eq.~(\ref{Pot}), $Y^{(k_Y)}_{I_Y}$ should transform as a multiplet modular form of weight $k_Y$ and representation $I_Y$, namely
\begin{eqnarray}
\gamma: ~ Y^{(k_Y)}_{I_Y} (\tau) \rightarrow Y^{(k_Y)}_{I_Y} (\gamma\tau) = (c\tau + d)^{k_Y} \rho^{}_{I_Y}(\gamma) Y^{(k_Y)}_{I_Y} (\tau) \;.
\label{MG:mf}
\end{eqnarray}
The weight $k_Y$ and representation $\rho^{}_{I_Y}$ must satisfy two conditions: $k_Y = k_{i_1} + \cdots + k_{i_n}$ holds and $\rho_{I_Y} \otimes \rho_{I_1} \otimes \cdots \otimes \rho_{I_n}$ contains at least one invariant singlet.

\def\thefootnote{\arabic{footnote}}
\setcounter{footnote}{0}
\setcounter{table}{0}
\setcounter{equation}{0}
\setcounter{figure}{0}

\section{Realizations of Two-zero Textures}
In the basis where the charged lepton mass matrix $M_\ell$ is diagonal, there are totally fifteen different two-zero textures of the Majorana neutrino mass matrix $M_\nu$~\cite{Frampton:2002yf,Xing:2002ta,Xing:2002ap}. Among them, eight cases have been ruled out by current neutrino oscillation data. Only seven two-zero textures are left compatible with current experimental data at $3\sigma$ level~\cite{Frampton:2002rn,Kageyama:2002zw,Desai:2002sz,Frigerio:2002fb,Kaneko:2002yp,Bhattacharyya:2002aq,Honda:2003pg,Dev:2006qe,Branco:2007nn,Grimus:2011sf,Fritzsch:2011qv,Meloni:2014yea,Cebola:2015dwa,Zhou:2015qua,Singh:2016qcf,Alcaide:2018vni,Singh:2019baq,Borgohain:2019pya}. They are usually labeled ${\bf A_{1,2}}$, ${\bf B_{1,2,3,4}}$ and ${\bf C}$, whose specific textures are
\begin{eqnarray}
&{\bf A_1}&: \left(\begin{matrix} 0 & 0 & \times \\ 0 & \times & \times \\ \times & \times & \times \end{matrix}\right) \;,\qquad {\bf A_2}: \left(\begin{matrix} 0 & \times & 0 \\ \times & \times & \times \\ 0 & \times & \times \end{matrix}\right) \;,
\nonumber
\\
&{\bf B_1}&: \left(\begin{matrix} \times & \times & 0 \\ \times & 0 & \times \\ 0 & \times & \times \end{matrix}\right) \;,\qquad {\bf B_2}: \left(\begin{matrix} \times & 0 & \times \\ 0 & \times & \times \\ \times & \times & 0 \end{matrix}\right) \;,
\nonumber
\\
&{\bf B_3}&: \left(\begin{matrix} \times & 0 & \times \\ 0 & 0 & \times \\ \times & \times & \times \end{matrix}\right) \;,\qquad {\bf B_4}: \left(\begin{matrix} \times & \times & 0 \\ \times & \times & \times \\ 0 & \times & 0 \end{matrix}\right) \;,
\nonumber
\\
&{\bf C}&: \left(\begin{matrix} \times & \times & \times \\ \times & 0 & \times \\ \times & \times & 0 \end{matrix}\right) \;,
\label{tz}
\end{eqnarray}
in which ``$\times$" represents a non-zero entry in the corresponding position. 

Two-zero textures of the Majorana neutrino mass matrix can be realized either with Abelian symmetry groups~\cite{Fritzsch:2011qv,Grimus:2004hf,Grimus:2004az,Dev:2011jc,Felipe:2014vka,Borgohain:2018lro} or with non-Abelian symmetry groups such as $A_4$~\cite{Zhou:2015qua,Hirsch:2007kh}. But in these flavor symmetry models, a lot of flavons have to be introduced and we must carefully deal with rather complicated potentials of theirs. To reduce the annoying degrees of freedom, we make use of the modular $A_4$ symmetry to achieve the seven allowed two-zero textures in this work, which is free from flavons. The reason for choosing the $A_4$ group is that $A_4$ is the only group who has three inequivalent one-dimensional representations among the non-Abelian discrete groups isomorphic to $\Gamma_N$. It is natural to assign three-generation leptons to different one-dimensional representations such that the diagonal charged lepton mass matrix can be obtained easily. In addition, not all modular forms of a specific weight and level 3 can be arranged into three inequivalent singlets of $A_4$ simultaneously, which makes it possible to provide enough degrees of freedom to gain zero entries in the Majorana neutrino mass matrix when all relevant particles are assigned as singlets~(${\bf 1}$, ${\bf 1^{\prime}}$ or ${\bf 1^{\prime\prime}}$) of $A_4$. The modular forms of weights up to 10 and level 3 forming singlets of $A_4$ are listed in Table~\ref{sing}.

\begin{table}[h]
	\centering
	\vspace{-0.2cm}
	\caption{The modular forms of weights up to 10 and level 3 which form singlets of $A_4$.}
	\setlength{\tabcolsep}{0.8cm}
	\vspace{0.2cm}
	\begin{tabular}{|c|l|}
		\hline
		${\bf 1}$ &  $Y^{(4)}_{\bf 1}$ ~, \quad $Y^{(6)}_{\bf 1}$ ~, \quad $Y^{(8)}_{\bf 1}$ ~, \quad $Y^{(10)}_{\bf 1}$
		\\
		\hline
		${\bf 1^\prime}$ &  $Y^{(4)}_{\bf 1^\prime}$ ~, \quad  $Y^{(8)}_{\bf 1^\prime}$ ~, \quad  $Y^{(10)}_{\bf 1^\prime}$ 
		\\
		\hline
		${\bf 1^{\prime\prime}}$ &  $Y^{(8)}_{\bf 1^{\prime\prime}}$ 
		\\
		\hline
	\end{tabular}
\label{sing}
\end{table}

In the present paper, we consider neutrino masses originating either from the Weinberg operator~\cite{Weinberg:1979sa} or form the type-\uppercase\expandafter{\romannumeral1} seesaw mechanism~\cite{Minkowski:1977sc,Yanagida:1979as,GellMann:1980vs,Glashow:1979nm,Mohapatra:1979ia}. For the former case, the superpotential in the lepton sector is 
\begin{eqnarray}
W = \alpha (E^cH_dLf_E(Y))_{\bf 1} + \frac{1}{\Lambda} (H_uH_uLLf_W(Y))_{\bf 1} \;,
\label{Wein}
\end{eqnarray}
and for the latter case, it reads
\begin{eqnarray}
W = \alpha (E^cH_dLf_E(Y))_{\bf 1} + g (N^cH_uLf_N(Y))_{\bf 1} + M (N^cN^cf_M(Y))_{\bf 1} \;,
\label{seesaw}
\end{eqnarray}
where $\alpha$, $\Lambda$, $g$, $M$ are constant coefficients and $f_{E,W,N,M}(Y)$ denote the modular form multiplets. The compact forms of the superpotential in Eqs.~(\ref{Wein}) and (\ref{seesaw}) may hide some arbitrary coefficients associated with independent invariant singlets arising from the same term. As to the K$\ddot{\rm a}$hler potential, we take the form given in Eq.~(\ref{Kp}). The chiral supermultiplets and a typical assignment of representations and weights are shown in Table \ref{cont}, where $k_i$, $r_i$ and $l_i$ ($i=1,2,3$) are modular weights of the corresponding chiral superfields, and $N^c_i$ only exist in the case where the type-\uppercase\expandafter{\romannumeral1} seesaw mechanism is applied to generate neutrino masses. And in some specific models, the representation and weight assignments for the chiral superfields given in Table \ref{cont} will be slightly modified.

\begin{table}[h]
	\centering
	\vspace{-0.2cm}
	\caption{A typical assignment of representations and weights for the chiral supermultiplets. The chiral superfields $N^c_1$, $N^c_2$, $N^c_3$ are only considered in the case where neutrino masses come from the type-\uppercase\expandafter{\romannumeral1} seesaw mechanism.}
	\setlength{\tabcolsep}{0.3cm}
	\vspace{0.2cm}
	\begin{tabular}{|c|c|c|c|c|c|c|c|c||c|c|c|}
		\hline
		& $L_1$ & $L_2$ & $L_3$ & $E^c_1$ & $E^c_2$ & $E^c_3$ & $H_u$ & $H_d$ & $N^c_1$ & $N^c_2$ & $N^c_3$\\
		\hline
		$A_4$ & ${\bf 1}$ &  ${\bf 1^\prime}$ &  ${\bf 1^{\prime\prime}}$ &  ${\bf 1}$ &  ${\bf 1^{\prime\prime}}$ &  ${\bf 1^\prime}$ &  ${\bf 1}$ &  ${\bf 1}$ & ${\bf 1}$ & ${\bf 1^\prime}$ & ${\bf 1^{\prime\prime}}$ \\
		\hline
		$-k$ & $k_1$ & $k_2$ & $k_3$ & $r_1$ & $r_2$ & $r_3$ & 0 & 0 & $l_1$ & $l_2$ & $l_3$ \\
		\hline
	\end{tabular}
	\label{cont}
\end{table}
\vspace{-0.3cm}

\subsection{Textures from the Weinberg Operator}
We first consider the case where neutrino masses are generated by the Weinberg operator. All seven allowed two-zero textures listed in Eq.~(\ref{tz}) can be achieved with different representation assignments and weights for the chiral superfields. We classify the seven allowed two-zero textures into three categories: ({\romannumeral1}) ${\bf B_{1,2}}$, (\romannumeral2) ${\bf A_{1,2}}$ and ${\bf B_{3,4}}$, (\romannumeral3) ${\bf C}$. If one texture is realized, other textures in the same category can be realized by exchanging or alternating the representations and weights of the corresponding chiral superfields. This feature can be seen from Eq.~(\ref{tz}) exactly. 

In the model-building process, we do not consider any modular forms of weights larger than 10, no matter whether they exist or not and we define the highest weight for modular forms involved in one model as $k_{Y{\rm max}} = \max\{-2k_i, -(k_m+r_n)\}$ with $i,m,n=1,2,3$. Usually more than one model can successfully achieve a specific two-zero texture, thus there exists a minimal $k_{Y{\rm max}}$. The minimal $k_{Y{\rm max}}$ for different cases is usually different, due to the special property of modular forms listed in Table~\ref{sing} and the different textures involved. In the following, we only consider the model whose $k_{Y{\rm max}}$ is minimal for a specific two-zero texture. All the possible models with larger $k_{Y{\rm max}}$ but not larger than 10 can be seen in Appendix B.1.

\begin{center}
(\romannumeral1) Textures ${\bf B_{1,2}}$
\end{center}

In this case, the minimal $k_{Y{\rm max}}$ is 4. We first assign representations for the chiral superfields as those shown in Table~\ref{cont} and the corresponding weights are
\begin{eqnarray}
k_i = -2 \;, \quad r_j = 2 
\label{case1:we}
\end{eqnarray}
with $i,j = 1, 2, 3$. The explicit form of the superpotential is 
\begin{eqnarray}
W = && \alpha E^c_1 H^{}_d L^{}_1 + \beta E^c_2 H^{}_d L^{}_2 + \gamma E^c_3 H^{}_d L^{}_3
\nonumber
\\
&&+ \frac{1}{\Lambda} H^{}_uH^{}_u \left[\lambda^{}_{11}Y^{(4)}_{\bf 1}L^{}_1L^{}_1 + \lambda^{}_{13}Y^{(4)}_{\bf 1^\prime}(L^{}_1L^{}_3 + L^{}_3L^{}_1) \right.
\nonumber
\\
&& \left. + \lambda^{}_{22} Y^{(4)}_{\bf 1^\prime} L^{}_2L^{}_2 + \lambda^{}_{23} Y^{(4)}_{\bf 1} (L^{}_2L^{}_3 + L^{}_3L^{}_2) \right] \;,
\label{case1:sp}
\end{eqnarray}
where $\alpha$, $\beta$, $\gamma$, $\lambda_{ij}~(i,j=1,2,3)$ are arbitrary coefficients, and it is obvious that no non-trivial modular form is involved in the charged lepton sector. Whereas for the model whose $k_{Y{\rm max}}$ is larger than 4, non-trivial modular forms are possible to be involved in the charged lepton sector, such as the model with $k_i = -2$, $r_j = -4~(i,j=1,2,3)$ and $k_{Y{\rm max}} = 6$. 

After spontaneous electroweak symmetry breaking, the superpotential in Eq.~(\ref{case1:sp}) leads to the following charged lepton mass matrix and Majorana neutrino mass matrix~\footnote{We work in the left-right convention for all lepton mass terms.}:
\begin{eqnarray}
M_\ell = v_d \left(\begin{matrix} \alpha^\ast & 0 & 0 \\ 0 & \beta^\ast & 0 \\0 & 0 & \gamma^\ast \end{matrix}\right) \;,\quad M_\nu = \frac{v^2_u}{\Lambda} \left(\begin{matrix} \lambda^\ast_{11} Y^{(4)\ast}_{\bf 1} & 0 & \lambda^\ast_{13} Y^{(4)\ast}_{\bf 1^\prime} \\ 0 & \lambda^\ast_{22} Y^{(4)\ast}_{\bf 1^\prime} & \lambda^\ast_{23} Y^{(4)\ast}_{\bf 1} \\ \lambda^\ast_{13} Y^{(4)\ast}_{\bf 1^\prime} &  \lambda^\ast_{23} Y^{(4)\ast}_{\bf 1} & 0 \end{matrix}\right) 
\label{case1:mm}
\end{eqnarray}
with $v_d = \langle H^0_d \rangle$ and $v_u = \langle H^0_u \rangle$. It is clear that the charged lepton mass matrix is diagonal and the Majorana mass matrix is exactly the texture labeled ${\bf B_2}$. 

The texture ${\bf B_1}$ can not be achieved with the representation assignments given in Table~\ref{cont}. But as mentioned above, the texture ${\bf B_1}$ can be obtained by exchanging the representation assignments and weights of $L_2$ and $L_3$ as well as those of $E^c_2$ and $E^c_3$ in Table~\ref{cont} and Eq.~(\ref{case1:we}), namely, exchanging $L_2 \leftrightarrow L_3$ and $E^c_2 \leftrightarrow E^c_3$ in the superpotential in Eq.~(\ref{case1:sp}). This treatment yields
\begin{eqnarray}
M_\ell = v_d \left(\begin{matrix} \alpha^\ast & 0 & 0 \\ 0 & \gamma^\ast & 0 \\0 & 0 & \beta^\ast \end{matrix}\right) \;,\quad 
M_\nu = \frac{v^2_u}{\Lambda} \left(\begin{matrix} \lambda^\ast_{11} Y^{(4)\ast}_{\bf 1} & \lambda^\ast_{13} Y^{(4)\ast}_{\bf 1^\prime} & 0  \\ \lambda^\ast_{13} Y^{(4)\ast}_{\bf 1^\prime} & 0 & \lambda^\ast_{23} Y^{(4)\ast}_{\bf 1} \\ 0 &  \lambda^\ast_{23} Y^{(4)\ast}_{\bf 1} & \lambda^\ast_{22} Y^{(4)\ast}_{\bf 1^\prime} \end{matrix}\right) \;,
\label{case1:mm1}
\end{eqnarray}
just corresponding to the texture ${\bf B_{1}}$. Doing the same thing for $L_1$ and $L_2$, $E^c_1$ and $E^c_2$ still gives the texture ${\bf B_2}$, and alternating the representation assignments and weights in the order, $L_1 \rightarrow L_3 \rightarrow L_2 \rightarrow L_1$ and $E^c_1 \rightarrow E^c_3 \rightarrow E^c_2 \rightarrow E^c_1$, also gives the texture ${\bf B_1}$.

\begin{center}
(\romannumeral2) Textures ${\bf A_{1,2}}$ and $\bf B_{3,4}$
\end{center}

In this case, the minimal $k_{Y{\rm max}}$ is 8 larger than that in the case (\romannumeral1). We first assume the representation assignments for the chiral superfields to be those in Table.~\ref{cont}, and the weights for the chiral superfields are
\begin{eqnarray}
k_1 = k_2 = -4 \;,\;  k_3 = -2 \;,\quad r_1 = 0 ~{\rm or}~ 4 \;,\; r_2 = -2 ~{\rm or}~ 4 \;,\; r_3 = -2 ~{\rm or}~ 2 \;,
\label{case2:we}
\end{eqnarray}
with which the superpotential is given by
\begin{eqnarray}
W = && \alpha Y^{(x)}_{\bf 1} E^c_1 H^{}_d L^{}_1 + \beta Y^{(y)}_{\bf 1} E^c_2 H^{}_d L^{}_2 + \gamma Y^{(z)}_{\bf 1} E^c_3 H^{}_d L^{}_3
\nonumber
\\
&&+ \frac{1}{\Lambda} H^{}_uH^{}_u \left[\lambda^{}_{11}Y^{(8)}_{\bf 1}L^{}_1L^{}_1 + \lambda^{}_{12}Y^{(8)}_{\bf 1^{\prime\prime}}(L^{}_1L^{}_2 + L^{}_2L^{}_1) \right.
\nonumber
\\
&& \left. + \lambda^{}_{22} Y^{(8)}_{\bf 1^\prime} L^{}_2L^{}_2 + \lambda^{}_{23} Y^{(6)}_{\bf 1} (L^{}_2L^{}_3 + L^{}_3L^{}_2) \right] \;,
\label{case2:sp}
\end{eqnarray}
where $x=0$ or $4$, $y=0$ or $6$ and $z=0$ or $4$ with $Y^{(0)}_{\bf 1}$ being constant and independent of $\tau$. In the charged lepton sector, non-trivial modular forms can get involved if $r_1=4$, $r_2=4$ and $r_3=2$ do not hold simultaneously. The charged lepton mass matrix remains diagonal no matter what values of $r_1$, $r_2$ and $r_3$ are taken. The charged lepton and Majorana neutrino mass matrices turn out to be
\begin{eqnarray}
M_\ell = v_d \left(\begin{matrix} \alpha^\ast Y^{(x)\ast}_{\bf 1} & 0 & 0 \\ 0 & \beta^\ast Y^{(y)\ast}_{\bf 1} & 0 \\ 0 & 0 & \gamma^{\ast} Y^{(z)\ast}_{\bf 1} \end{matrix}\right) \;,\quad
M_\nu = \frac{v^2_u}{\Lambda} \left(\begin{matrix} \lambda^\ast_{11} Y^{(8)\ast}_{\bf 1} & \lambda^\ast_{12} Y^{(8)\ast}_{\bf 1^{\prime\prime}} & 0 \\ \lambda^\ast_{12} Y^{(8)\ast}_{\bf 1^{\prime\prime}} & \lambda^\ast_{22} Y^{(8)\ast}_{\bf 1^\prime} & \lambda^\ast_{23} Y^{(6)\ast}_{\bf 1} \\ 0 &  \lambda^\ast_{23} Y^{(6)\ast}_{\bf 1} & 0 \end{matrix}\right) 
\label{case2:mm}
\end{eqnarray}
which coincide with the texture ${\bf B_4}$. 

Similar to the case (\romannumeral1), exchanging the representation assignments and weights of $L_2$ and $L_3$ together with those of $E^c_2$ and $E^c_3$ in Table~\ref{cont} and Eq.~(\ref{case2:we}) leads to the texture ${\bf B_{3}}$. The texture ${\bf A_2}$ can be obtained by exchanging the representation assignments and weights of $L_1$ and $L_3$ and also those of $E^c_1$ and $E^c_3$. Note that exchanges between any two chiral superfields like before can not give the texture ${\bf A_1}$. Instead, it needs to alternate the representation assignments and weights of $L_i$ and those of $E^c_i$ with $i = 1, 2, 3$. More specifically, alternating the representation assignments and weights in the order, $L_1 \rightarrow L_2 \rightarrow L_3 \rightarrow L_1$ and $E^c_1 \rightarrow E^c_2 \rightarrow E^c_3 \rightarrow E^c_1$, results in the texture ${\bf A_1}$. In fact, exchanging or alternating the representation assignments and weights of the chiral superfields corresponds to the permutations of mass matrices which are just the relations between the textures in one category.

\begin{center}
(\romannumeral3) Texture ${\bf C}$ 
\end{center}

The minimal $k_{Y{\rm max}}$ in this case is 10. Taking the representation assignments in Table~\ref{cont}, there are two totally different choices of the weights for the chiral superfields. One of them is 
\begin{eqnarray}
k_1 = -5 \;,\; k_2 = k_3 = -3 \;,\quad r_1 = -3, 1 ~{\rm or}~ 5 \;,\; r_2 = -1 ~{\rm or}~ 3 \;,\; r_3 = 3 \;,
\label{case3:we1}
\end{eqnarray}
and the other is given by 
\begin{eqnarray}
k_1 = k_3 = -5 \;,\; k_2 = -3 \;,\quad r_1 = -1 ~{\rm or}~ 5 \;,\; r_2 = -1 ~{\rm or}~ 3 \;,\; r_3 = 1  ~{\rm or}~ 5 \;.
\label{case3:we2}
\end{eqnarray}
Then the superpotential reads:
\begin{eqnarray}
W = && \alpha Y^{(x^\prime)}_{\bf 1} E^c_1 H^{}_d L^{}_1 + \beta Y^{(y\prime)}_{\bf 1} E^c_2 H^{}_d L^{}_2 + \gamma Y^{(z^\prime)}_{\bf 1} E^c_3 H^{}_d L^{}_3
\nonumber
\\
&&+ \frac{1}{\Lambda} H^{}_uH^{}_u \left[\lambda^{}_{11}Y^{(10)}_{\bf 1}L^{}_1L^{}_1 + \lambda^{}_{12}Y^{(8)}_{\bf 1^{\prime\prime}}(L^{}_1L^{}_2 + L^{}_2L^{}_1) \right.
\nonumber
\\
&& \left. + \lambda^{}_{13} Y^{(p)}_{\bf 1^\prime} (L^{}_1L^{}_3 + L^{}_3L^{}_1) + \lambda^{}_{23} Y^{(q)}_{\bf 1} (L^{}_2L^{}_3 + L^{}_3L^{}_2) \right] \;,
\label{case3:sp}
\end{eqnarray}
where $x^\prime=0,4$ or $8$, $y^\prime=0$ or $4$, $z^\prime=0$, $p=8$ and $q=6$ for the weights in Eq.~(\ref{case3:we1}), and $x^\prime=0$ or $6$, $y^\prime=0$ or $4$, $z^\prime=0$ or $4$, $p=10$ and $q=8$ for the weights in Eq.~(\ref{case3:we2}). No matter whether non-trivial modular forms have a hand in the charged lepton sector or not, the charged lepton mass matrix remains diagonal. Then the charged lepton and Majorana mass matrices are given by
\begin{eqnarray}
M_\ell = v_d \left(\begin{matrix} \alpha^\ast Y^{(x^\prime)\ast}_{\bf 1} & 0 & 0 \\ 0 & \beta^\ast Y^{(y^\prime)\ast}_{\bf 1} & 0 \\ 0 & 0 & \gamma^{\ast} Y^{(z^\prime)\ast}_{\bf 1} \end{matrix}\right) \;,\quad
M_\nu = \frac{v^2_u}{\Lambda} \left(\begin{matrix} \lambda^\ast_{11} Y^{(8)\ast}_{\bf 1} & \lambda^\ast_{12} Y^{(8)\ast}_{\bf 1^{\prime\prime}} &  \lambda^\ast_{13} Y^{(p)\ast}_{\bf 1^{\prime}} \\ \lambda^\ast_{12} Y^{(8)\ast}_{\bf 1^{\prime\prime}} & 0 & \lambda^\ast_{23} Y^{(q)\ast}_{\bf 1} \\ \lambda^\ast_{13} Y^{(p)\ast}_{\bf 1^{\prime}}  &  \lambda^\ast_{23} Y^{(q)\ast}_{\bf 1} & 0 \end{matrix}\right) \;.
\label{case3:mm}
\end{eqnarray}
So we arrive at the texture ${\bf C}$ exactly. The texture ${\bf C}$ can also be obtained if we exchange the representation assignments and weights of $L_2$ and $L_3$ and those of $E^c_2$ and $E^c_3$ in Table~\ref{cont} and Eq.~(\ref{case3:we1}) or Eq.~(\ref{case3:we2}).

\subsection{Textures from the Seesaw Mechanism}
Embedded into the tpye-\uppercase\expandafter{\romannumeral1} seesaw mechanism where three right-handed neutrinos are introduced, the two-zero textures can be also realized. Due to three new degrees of freedom, the situation becomes more complicated than the case where neutrino masses originate from the Weinberg operator. For simplicity, we assume that the representation assignments of three right-handed neutrinos are fixed as given in Table~\ref{cont}. In addition, we fix the mass matrix of right-handed neutrinos to the following form: 
\begin{eqnarray}
M_{\rm R} = \left(\begin{matrix} a & 0 & 0 \\ 0 & 0 & b \\ 0 & b & 0 \end{matrix} \right) \;,
\label{rm}
\end{eqnarray}
which can be easily achieved with the fixed representation assignments. Then with the seesaw formula $M^{}_\nu = - M^{}_{\rm D} M^{-1}_{\rm R} M^{T}_{\rm D}$ and Eq.~(\ref{rm}), the zero entries in the effective neutrino mass matrix $M_\nu$ result from zero textures of the Dirac neutrino mass matrix~\footnote{We do not consider the cases where special relations exist among the elements of $M_{\rm D}$.}. Owing to the intrinsic structure of $M_{\rm R}$ given in Eq.~(\ref{rm}), the textures ${\bf B_{1,2}}$ can not be generated from any zero textures of $M_{\rm D}$. All the textures of $M_{\rm D}$ which lead to the textures ${\bf A_{1,2}}$, ${\bf B_{3,4}}$ and ${\bf C}$ are listed in Appendix B. Moreover, only four-zero and five-zero textures of $M_{\rm D}$ can be realized with the modular $A_4$ symmetry in our framework. 

Similarly to section 3.1, we define the highest weight for modular forms in one model as $k^{\prime}_{Y{\rm max}} = \max{\{ -2l_i, -(k_m + l_n), -(k_g + r_f)\}} $ with $i,m,n,g,f = 1, 2, 3$. Here we only consider four-zero textures of $M_{\rm D}$ and then the models with minimal $k^{\prime}_{Y{\rm max}}$. The information for all other cases up to $k^{\prime}_{Y{\rm max}} = 10$ is shown in Appendix B.2.

\begin{center}
(a) Textures ${\bf B_{3,4}}$ and ${\bf A_{1,2}}$
\end{center}

First, we consider the representation assignments exactly given in Table~\ref{cont} and the following weights:
\begin{eqnarray}
k_1=k_3=-4\;,\; k_2=-6\;, \quad l_1=l_2=l_3=0\;, \quad r_1 = 4\;,\; r_2 = 2 ~{\rm or}~ 6\;,\; r_3 = 0 ~{\rm or}~ 4\;,
\label{casea:we}
\end{eqnarray}
with $k^{\prime}_{Y{\rm max}} = 6$. Then the superpotential can be written in a explicit form:
\begin{eqnarray}
W = &&\alpha E^c_1H^{}_dL^{}_1 + \beta Y^{(x^{\prime\prime})}_{\bf 1} E^c_2H^{}_dL^{}_2 + \gamma Y^{(y^{\prime\prime})}_{\bf 1} E^c_3H^{}_dL^{}_3
\nonumber
\\
&& + \lambda_{11} Y^{(4)}_{\bf 1} N^c_1H^{}_uL^{}_1 + \lambda_{13} Y^{(4)}_{\bf 1^{\prime}} N^c_1H^{}_uL_3 + \lambda_{23} Y^{(4)}_{\bf 1} N^c_2H^{}_uL^{}_3 
\nonumber
\\
&&+ \lambda_{31} Y^{(4)}_{\bf 1^\prime} N^c_3H^{}_uL^{}_1 + \lambda_{32} Y^{(6)}_{\bf 1} N^c_3H^{}_uL^{}_2
\nonumber
\\
&& + M^{}_{11} N^c_1N^c_1 + M^{}_{23} (N^c_2N^c_3 + N^c_3N^c_2) \;,
\label{casea:sp}
\end{eqnarray}
where $x^{\prime\prime}, y^{\prime\prime} =0, 4$. After spontaneous electroweak symmetry breaking, the mass matrices read
\begin{eqnarray}
M_\ell = v_d \left(\begin{matrix} \alpha^\ast & 0 & 0 \\ 0 & \beta^\ast Y^{(x^{\prime\prime})\ast}_{\bf 1} & 0 \\ 0 & 0 & \gamma^{\ast} Y^{(y^{\prime\prime})\ast}_{\bf 1} \end{matrix}\right) 
\label{casea:ml}
\end{eqnarray}
for charged leptons which is diagonal and 
\begin{eqnarray} 
M_{\rm D} = v_u \left(\begin{matrix} \lambda^{\ast}_{11}Y^{(4)\ast}_{\bf 1} & 0 & \lambda^{\ast}_{31} Y^{(4)\ast}_{\bf 1^\prime} \\ 0 & 0 & \lambda^{\ast}_{32} Y^{(6)\ast}_{\bf 1} \\ \lambda^{\ast}_{13} Y^{(4)\ast}_{\bf 1^\prime} & \lambda^{\ast}_{23} Y^{(4)\ast}_{\bf 1} & 0 \end{matrix}\right) \;,\quad M_{\rm R} = \left(\begin{matrix} M^{\ast}_{11} & 0 & 0 \\ 0 & 0 & M^{\ast}_{23} \\ 0 & M^{\ast}_{23} & 0\end{matrix}\right) \;,
\label{casea:mdmr}
\end{eqnarray}
for neutrinos, where $M_{\rm R}$ has the structure given in Eq.~(\ref{rm}) and $M_{\rm D}$ has four zero entries. With the help of the seesaw formula, the effective neutrino mass matrix turns out to be 
\begin{eqnarray}
M_\nu = -M^{}_{\rm D}M^{-1}_{\rm R}M^{T}_{\rm D} = \left(\begin{matrix} A & 0 & B \\ 0 & 0 & C \\ B & C & D\end{matrix}\right)
\label{casea:mv}
\end{eqnarray}
with
\begin{eqnarray}
A &=& - \frac{v^2_u \lambda^{2\ast}_{11} Y^{(4)2\ast}_{\bf 1}}{M^{\ast}_{11}} \;,\quad
B = - \frac{v^2_u\lambda^{\ast}_{11} \lambda^{\ast}_{13} Y^{(4)\ast}_{\bf 1} Y^{(4)\ast}_{\bf 1^\prime}}{M^{\ast}_{11}} - \frac{v^2_u\lambda^{\ast}_{23} \lambda^{\ast}_{31}Y^{(4)\ast}_{\bf 1} Y^{(4)\ast}_{\bf 1^\prime}}{M^{\ast}_{23}} \;,
\nonumber
\\
C &=& - \frac{v^2_u\lambda^{\ast}_{23} \lambda^{\ast}_{32}Y^{(4)\ast}_{\bf 1} Y^{(6)\ast}_{\bf 1}}{M^{\ast}_{23}} \;,\quad
D = - \frac{v^2_u\lambda^{2\ast}_{13} Y^{(4)2\ast}_{\bf 1^\prime}}{M^{\ast}_{11}} \;,
\label{casea:mvp}
\end{eqnarray}
which is exactly the texture ${\bf B_3}$.

The textures ${\bf B_4}$ and ${\bf A_{1,2}}$ can not be realized with $k^{\prime}_{Y{\rm max}} = 6$ and the representation assignments given in Table~\ref{cont}, as can be seen in Appendix B.2. Nonetheless, if we keep the representation and weight assignments of the right-handed neutrinos unchanged and exchange the representation assignments and weights of $L_2$ and $L_3$ together with those of $E^c_2$ and $E^c_3$ given in Table~\ref{cont} and Eq.~(\ref{casea:we}), we can obtain the texture ${\bf B_4}$. Similarly, exchanging the representation assignments and weights of $L_1$ and $L_2$ and those of $E^c_1$ and $E^c_2$ gives the texture ${\bf A_1}$. The texture ${\bf A_2}$ can be achieved by alternating the representation assignments and weights of $L_i$ and $E^c_j$ in the order, $L_1 \rightarrow L_3 \rightarrow L_2 \rightarrow L_1$ and $E^c_1 \rightarrow E^c_3 \rightarrow E^c_2 \rightarrow E^c_1$.

\begin{center}
(b) Texture ${\bf C}$ 
\end{center}

For the texture ${\bf C}$, the minimal $k^{\prime}_{Y{\rm max}}$ is 6 with a five-zero texture of $M_{\rm D}$ which can be seen from Appendix B.2. But here we only consider four-zero textures of $M_{\rm D}$ as mentioned above. In this case, the minimal $k^{\prime}_{Y{\rm max}} $ is 8. The representation assignments are given in Table~\ref{cont} and the weights are
\begin{eqnarray}
k_1 = -8\;,\; k_2 = k_3 =-6\;,\quad l_1=l_2=l_3=0\;,\quad r_1= 0, 4 ~{\rm or}~ 8\;,\; r_2=2 ~{\rm or}~ 6 \;,\; r_3 = 6\;,
\label{caseb:we1}
\end{eqnarray}
or
\begin{eqnarray}
k_1 = -5\;,\; k_2 = k_3 =-3\;,\quad l_1=l_2=l_3=-3\;,\quad r_1= -3, 1 ~{\rm or}~ 5\;,\; r_2=-1 ~{\rm or}~ 3 \;,\; r_3 = 3\;.\quad
\label{caseb:we2}
\end{eqnarray}
Then the superpotential is given by
\begin{eqnarray}
W =&& \alpha Y^{(x^{\prime\prime\prime})}_{\bf 1} E^c_1H^{}_dL^{}_1 + \beta Y^{(y^{\prime\prime\prime})}_{\bf 1} E^c_2H^{}_dL^{}_2 + \gamma E^c_3H^{}_dL^{}_3
\nonumber
\\ 
&& + \lambda^{}_{11} Y^{(8)}_{\bf 1} N^c_1H^{}_uL^{}_{1} + \lambda^{}_{21} Y^{(8)}_{\bf 1^{\prime\prime}} N^c_2H^{}_uL^{}_1 + \lambda^{}_{23} Y^{(6)}_{\bf 1} N^c_2H^{}_uL^{}_3
\nonumber
\\
&& + \lambda^{}_{31} Y^{(8)}_{\bf 1^\prime} N^c_3H^{}_uL^{}_1 + \lambda^{}_{32} Y^{(6)}_{\bf 1} N^c_3H^{}_uL^{}_2
\nonumber
\\
&& + M^{}_{11} Y^{(p^\prime)}_{\bf 1} N^c_1N^c_1 + M^{}_{23} Y^{(p^\prime)}_{\bf 1} (N^c_2N^c_3 + N^c_3N^c_2) \;,
\label{caseb:sp}
\end{eqnarray}
with $x^{\prime\prime\prime}=0,4,8$, $y^{\prime\prime\prime} = 0,4$, $p^\prime = 0$ for the weights in Eq.~(\ref{caseb:we1}), and $x^{\prime\prime\prime}=0,4,8$, $y^{\prime\prime\prime} = 0,4$, $p^\prime = 6$ for the weights in Eq.~(\ref{caseb:we2}). The above superpotential gives the following charged lepton and neutrino mass matrices:
\begin{eqnarray}
M_\ell = v_d \left(\begin{matrix} \alpha^\ast Y^{(x^{\prime\prime\prime})\ast}_{\bf 1} & 0 & 0 \\ 0 & \beta^\ast Y^{(y^{\prime\prime\prime})\ast}_{\bf 1} & 0 \\ 0 & 0 & \gamma^\ast \end{matrix}\right) \;,
\label{caseb:ml}
\end{eqnarray}
and 
\begin{eqnarray}
M_{\rm D} = v_u \left(\begin{matrix} \lambda^{\ast}_{11} Y^{(8)\ast}_{\bf 1} & \lambda^{\ast}_{21} Y^{(8)\ast}_{\bf 1^{\prime\prime}} & \lambda^{\ast}_{31} Y^{(8)\ast}_{\bf 1^\prime} \\ 0 & 0 & \lambda^{\ast}_{32} Y^{(6)\ast}_{\bf 1} \\ 0 & \lambda^{\ast}_{23} Y^{(6)\ast}_{\bf 1} & 0 \end{matrix}\right) \;,\quad M_{\rm R} = \left(\begin{matrix} M^{\ast}_{11} Y^{(p^{\prime})\ast}_{\bf 1} & 0 & 0 \\ 0 & 0 & M^{\ast}_{23} Y^{(p^\prime)\ast}_{\bf 1} \\ 0 & M^{\ast}_{23} Y^{(p^\prime)\ast}_{\bf 1} & 0 \end{matrix}\right) \;.
\label{caseb:mdmr}
\end{eqnarray}
With the seesaw formula, we can obtain the effective neutrino mass matrix:
\begin{eqnarray}
M_\nu = -M^{}_{\rm D}M^{-1}_{\rm R}M^{T}_{\rm D} =  \left(\begin{matrix} A^\prime & B^\prime & C^\prime \\ B^\prime & 0 & D^\prime \\ C^\prime & D^\prime & 0 \end{matrix}\right) \;,
\label{caseb:mv}
\end{eqnarray}
where
\begin{eqnarray}
A^\prime &=& - \frac{v^2_u\lambda^{2\ast}_{11} Y^{(8)2\ast}_{\bf 1}}{M^{\ast}_{11} Y^{(p^\prime)\ast}_{\bf 1}} - 2 \frac{v^2_u\lambda^{\ast}_{21} \lambda^{\ast}_{31} Y^{(8)\ast}_{\bf 1^{\prime}} Y^{(8)\ast}_{\bf 1^{\prime\prime}}}{ M^{\ast}_{23} Y^{(p^\prime)\ast}_{\bf 1} } \;,\quad B^\prime = - \frac{v^2_u\lambda^{\ast}_{21} \lambda^{\ast}_{32} Y^{(6)\ast}_{\bf 1} Y^{(8)\ast}_{\bf 1^{\prime\prime}}}{ M^{\ast}_{23} Y^{(p^\prime)\ast}_{\bf 1} } \;,
\nonumber
\\
C^\prime &=& - \frac{v^2_u\lambda^{\ast}_{23} \lambda^{\ast}_{31} Y^{(6)\ast}_{\bf 1} Y^{(8)\ast}_{\bf 1^{\prime}}}{ M^{\ast}_{23} Y^{(p^\prime)\ast}_{\bf 1} } \;,\quad D^\prime = -\frac{v^2_u\lambda^{\ast}_{23} \lambda^{\ast}_{32} Y^{(6)2\ast}_{\bf 1}}{ M^{\ast}_{23} Y^{(p^\prime)\ast}_{\bf 1} } \;.
\label{caseb:mvp}
\end{eqnarray}
The effective neutrino mass matrix in Eq.~(\ref{caseb:mv}) is just the texture ${\bf C}$. If the representation assignments and weights of right-handed neutrinos keep unchanged and those of $L_2$ and $L_3$ as well as of $E^c_2$ and $E^c_3$ are exchanged, it also leads us to the texture ${\bf C}$.
\vspace{0.2cm}

One can see that two-zero textures of the Majorana neutrino mass matrix can be realized in the modular $A_4$ invariant models without flavons, where neutrino masses originate either from the Weinberg operator or from the type-\uppercase\expandafter{\romannumeral1} seesaw mechanism. However for the latter case, thanks to the intrinsic structure of the right-handed neutrino mass matrix in our assumption for simplicity, only five textures can be achieved. Some comments on the results are in order.
\begin{itemize}
	\item It is easy to see that all the models that we have considered can not give any new constraints on two-zero textures of the Majorana neutrino mass matrix, which means that these results can not give any new phenomenological predictions compared to previous works. Since the phenomenological consequences of these two-zero textures have been systematically studied~\cite{Frampton:2002yf,Xing:2002ta,Xing:2002ap,Guo:2002ei,Xing:2003ez,Xing:2019vks,Frampton:2002rn,Kageyama:2002zw,Desai:2002sz,Frigerio:2002fb,Kaneko:2002yp,Bhattacharyya:2002aq,Honda:2003pg,Dev:2006qe,Branco:2007nn,Grimus:2011sf,Fritzsch:2011qv,Meloni:2014yea,Cebola:2015dwa,Zhou:2015qua,Singh:2016qcf,Alcaide:2018vni,Singh:2019baq,Borgohain:2019pya}, there is no need to discuss these issues repeatedly in this work.
	\item After the modulus $\tau$ acquires its VEV $\langle \tau \rangle$, the modular $A_4$ symmetry is broken. In general, the modular forms given in Table~\ref{sing} which form the three inequivalent singlets of $\Gamma_N$ are all not zero at $\langle \tau \rangle$. However, in some special cases, such as at the stabilizers, some modular forms that we have considered can be zero. This can make the zero entries in the Majorana neutrino mass matrix increase, that is to say, not all $\langle \tau \rangle$s can guarantee the obtained two-zero textures of the Majorana neutrino mass matrix. For example, at the stabilizer $\langle \tau \rangle = \tau_L = -1/2 + i \sqrt{3}/2$~\cite{Novichkov:2018yse,Gui-JunDing:2019wap}, $Y^{(4)}_{\bf 1} = Y^{(8)}_{\bf 1} = Y^{(8)}_{\bf 1^\prime} = Y^{(10)}_{\bf 1} = 0$ holds, so all the models involving these four modular forms can no longer give two-zero textures of the Majorana neutrino mass matrix.
	\item It is obvious that for models realizing different two-zero textures of $M_\nu$, the minimal $k^{}_{Y{\rm max}}$ or $k^{\prime}_{Y{\rm max}}$ is different. As seen in appendix B, the modular weights of modular forms need to achieve 10 so as to successfully realize all seven two-zero textures of $M_\nu$ in the case where neutrino masses originate from the Weinberg operator. For the textures with minimal $k_{Y{\rm max}}$ smaller than 10, there are more choices to achieve the textures. While in the case where the type-\uppercase\expandafter{\romannumeral1} seesaw mechanism is introduced to generate neutrino masses, the modular weights of modular forms only need to reach 6 to give five of the two-zero textures of $M_\nu$. The reason is that there are more degrees of freedom induced by three right-handed neutrinos compared to the former case. If we loose the limit on the modular weights involved, there will be more different models to realize two-zero textures of $M_{\nu}$.
	\item Actually, to successfully realize two-zero textures of $M_\nu$ in our framework, it is very essential that not all modular forms of a specific weight and level 3 can form three inequivalent singlets of $A_4$ simultaneously as shown in Table~\ref{sing}. This property makes it possible to gain zero entries in the Majorana neutrino mass matrix and keep charged lepton mass matrix diagonal at the same time. In principle, this approach can be applied to the Dirac neutrino case or the quark sector to achieve some interesting zero textures.
\end{itemize}

\section{Summary and Remarks} 
In this work, we have made the first attempt to realize the phenomenology favored two-zero textures of the Majorana neutrino mass matrix $M_\nu$ with the modular symmetry and without flavons. We have considered modular $A_4$ invariant models and derived the modular forms of weights up to 10 and level 3 which are holomorphic functions of the modulus $\tau$ and can be arranged into representations of the finite modular group $\Gamma_4 \simeq A_4$ as shown in Eqs.~(\ref{MF:2})---(\ref{MF:10}). To keep the models invariant under modular transformations, the chiral superfields and coupling constants or the right-handed neutrino mass matrix in the type-\uppercase\expandafter{\romannumeral1} seesaw mechanism which are regarded as modular forms must transform in the way given by Eqs.~(\ref{MG:csft}) and (\ref{MG:mf}). The representations and weights of the chiral superfields and modular forms have to satisfy the conditions that the relation $k_Y = k_{i_1} + \cdots + k_{i_n}$ holds and $\rho_{I_Y} \otimes \rho_{I_1} \otimes \cdots \otimes \rho_{I_n}$ contains at least one invariant singlet. We assign all chiral superfields into the three inequivalent singlets, ${\bf 1}$, ${\bf 1^\prime}$ and ${\bf 1^{\prime\prime}}$, of the $A_4$ group. Since not all modular forms of a specific weight and level 3 can be arranged into three inequivalent singlets of $A_4$ simultaneously as shown in Table~\ref{sing}, the charged lepton mass matrix can be made diagonal easily and the zero entries of $M_\nu$ can be achieved by properly assigning the representations and modular weights of the chiral superfields. Notice that arranging only three inequivalent singlets of the $A_4$ group is equivalent to considering a $Z_3$ symmetry and to some extent, the application of the modular weights is phenomenologically similar to imposing an additional Abelian symmetry, but it is not a real symmetry group and it depends on the properties and structures of the modular forms.

We have considered two cases where neutrino masses originate from the Weinberg operator and the type-\uppercase\expandafter{\romannumeral1} seesaw mechanism, respectively. In these two cases, we have defined $k^{}_{Y{\rm max}}$ and $k^{\prime}_{Y{\rm max}}$, respectively, which stand for the {\rm max}imal weight involved in a specific model. For a given two-zero texture, usually several models with different $k^{}_{Y{\rm max}}$ or $k^{\prime}_{Y{\rm max}}$ can realize the structure. Thus we only discuss models with minimal $k^{}_{Y{\rm max}}$ or $k^{\prime}_{Y{\rm max}}$ in detail. In the case where neutrino masses come from the Weinberg operator, all seven allowed two-zero textures of $M_\nu$ can be realized successfully, and the models with minimal $k_{Y{\rm max}}$ have been discussed in Section~3.1.  
In the case where neutrino masses originate from the type-\uppercase\expandafter{\romannumeral1} seesaw mechanism, for simplicity, we fix the representation assignments and the mass matrix $M_{\rm R}$ of three right-handed neutrinos as shown in Table~\ref{cont} and Eq.~(\ref{rm}). Then zero entries in $M_\nu$ result from zero textures of the Dirac neutrino mass matrix $M_{\rm D}$ via the seesaw formula $M^{}_\nu = - M^{}_{\rm D} M^{-1}_{\rm R} M^{T}_{\rm D}$. Due to the intrinsic structure of $M_{\rm R}$ in Eq.~(\ref{rm}), no matter whether flavor symmetries are imposed or not, only the textures ${\bf A_{1,2}}$, ${\bf B_{3,4}}$ and ${\bf C}$ can be achieved. For a specific two-zero texture of $M_\nu$, there are actually many textures of $M_{\rm D}$ and models to achieve it. We only consider four-zero textures of $M_{\rm D}$ and then the models with minimal $k^{\prime}_{Y{\rm max}}$. All the other possible models realizing two-zero textures of $M_\nu$ and the relations between these models have been discussed and shown in Appendix B. 
However, we can not obtain any new constraints to enhance the phenomenological predictive power of two-zero textures in these models and in some special case, the VEV of the modulus $\tau$ will destroy the obtained two-zero textures. 

We remark that it is interesting and instructive to realize two-zero textures of the Majorana neutrino mass matrix by imposing modular symmetries since there is no need to introduce a large number of flavons and it correlates approaches of the flavor symmetry and texture zeros to get an insight into the lepton flavor problem. It inspires us to explore modular symmetries of different levels and take different representation assignments for the chiral superfields in order to find some new constraints to enhance the phenomenological predictive power of two-zero textures of $M_\nu$. In principle, this approach can be applied to the Dirac neutrino case or the quark sector to achieve some fascinating zero textures.

\section*{Acknowledgments}
I am greatly indebted to Zhi-zhong Xing for his encouragements, carefully reading this manuscript and giving many useful suggestions. I would like to thank Xin Wang for helpful discussions. This work was supported in part by the National Natural Science Foundation of China under Grants No. 11775231 and No. 11835013.

\begin{appendices}
\setcounter{table}{0}
\setcounter{equation}{0}

\section{The $A_4$ Group}
The $A_4$ group is denoted as all even permutations of four objects  and it is the symmetry of a tetrahedron such that it contains twelve elements, which can be generated by two elements denoted as $S$ and $T$. These two generators satisfy the relations
\begin{eqnarray}
S^2 = (ST)^3 = T^3 = \mathbbm{1} \;. 
\label{Agen}
\end{eqnarray}
The twelve elements can be divided into four conjugate classes, so the $A_4$ group admits four irreducible representations --- three inequivalent one-dimensional representations, ${\bf 1}$, $ {\bf 1}^\prime  $, ${\bf 1^{\prime\prime}} $ and one three-dimensional representation, ${\bf 3}$. The irreducible representations of the $A_4$ group in a specific basis is given in Table~\ref{rep}, where $\omega = e^{2\pi i/3}$. With the representations shown in Table~\ref{rep}, the decomposition of tensor products is
\begin{eqnarray}
&{\bf 1} \otimes {\bf r} = {\bf r} \otimes {\bf 1} =  {\bf r} ~ ( {\bf r} =  {\bf 1}, {\bf 1^\prime}, {\bf 1^{\prime\prime}}, {\bf 3}) \;, &
\nonumber
\\
&{\bf 1^\prime} \otimes  {\bf 1^\prime} =  {\bf 1^{\prime\prime}} \;,\quad 
{\bf 1^{\prime\prime}} \otimes   {\bf 1^{\prime\prime}} = {\bf 1^\prime} \;,\quad 
{\bf 1^\prime} \otimes  {\bf 1^{\prime\prime}} = {\bf 1^{\prime\prime}} \otimes  {\bf 1^\prime} =  {\bf 1} \;, &
\nonumber
\\
&(\alpha)_{\bf 1} \otimes \left(\begin{matrix} a_1 \\ a_2 \\ a_3 \end{matrix}\right)_{\bf 3} =  \left(\begin{matrix}  \alpha a_1 \\ \alpha a_2 \\ \alpha a_3 \end{matrix}\right)_{\bf 3} \;,\quad 
(\beta)_{\bf 1^\prime} \otimes \left(\begin{matrix} a_1 \\ a_2 \\ a_3 \end{matrix}\right)_{\bf 3} =  \left(\begin{matrix}  \beta a_3 \\ \beta a_1 \\ \beta a_2 \end{matrix}\right)_{\bf 3} \;,\quad
(\gamma)_{\bf 1^{\prime\prime}} \otimes \left(\begin{matrix} a_1 \\ a_2 \\ a_3 \end{matrix}\right)_{\bf 3} =  \left(\begin{matrix}  \gamma a_2 \\ \gamma a_3 \\ \gamma a_1 \end{matrix}\right)_{\bf 3} \;,\quad & 
\nonumber
\\
&\left(\begin{matrix} a_1 \\ a_2 \\ a_3 \end{matrix}\right)_{\bf 3} \otimes \left(\begin{matrix} b_1 \\ b_2 \\ b_3 \end{matrix}\right)_{\bf 3} = \left(a_1b_1 + a_2b_3 + a_3b_2 \right)_{\bf 1} \oplus \left( a_3b_3 + a_1b_2 + a_2b_1 \right)_{\bf 1^\prime} \oplus \left( a_2b_2 + a_1b_3 + a_3b_1 \right)_{\bf 1^{\prime\prime}} &
\nonumber
\\
& \oplus \displaystyle \frac{1}{3} \left(\begin{matrix} 2a_1b_1 - a_2b_3 - a_3b_2 \\ 2a_3b_3 -a_1b_2 - a_2b_1 \\ 2a_2b_2 -a_1b_3 - a_3b_1 \end{matrix}\right)_{\bf 3} \oplus \displaystyle \frac{1}{2} \left(\begin{matrix}  a_2b_3 - a_3b_2 \\ a_1b_2 - a_2b_1 \\ a_1b_3 - a_3b_1 \end{matrix}\right)_{\bf 3} \;.&
\end{eqnarray}
Note that the decomposition of tensor products is unique, while the specific forms of reduced tensors are dependent on the representation of the $A_4$ group, which can be seen from refs.~\cite{Altarelli:2010gt,Ishimori:2010au,King:2013eh}.
\begin{table}[hpt]
	\centering
	\vspace{-0.2cm}
	\caption{The irreducible representations of the $A_4$ group where $\omega=e^{2\pi i/3}$.}
	\vspace{0.2cm}
	\begin{tabular}{|l|c|c|}
		\hline
		& $\rho(S)$ & $\rho(T)$ \\
		\hline
		${\bf 1}$ & 1 & 1 \\
		\hline
		${\bf 1^{\prime}}$ & 1 & $\omega$ \\
		\hline
		${\bf 1^{\prime\prime}}$ & 1 & $\omega^2$ \\
		\hline
		${\bf 3}$ & $\displaystyle \frac{1}{3} \left( \begin{matrix}
		-1 & 2 & 2 \\ 2 & -1 & 2 \\ 2 & 2 & -1 
		\end{matrix} \right)$ & $\displaystyle \left(\begin{matrix}
		1 & 0 & 0 \\ 0 & \omega & 0 \\ 0 & 0 & \omega^2
		\end{matrix}\right) $ \\
		\hline
	\end{tabular}
	\label{rep}
\end{table}
\vspace{-0.5cm}

\setcounter{table}{0}
\setcounter{equation}{0}
\section{All Models Achieving Two-zero Textures}
\subsection{The Weinberg Operator}
In Section 3.1, with neutrino masses originating from the Weinberg operator, we only give the model with minimal $k_{Y{\rm max}}$ in each case. Here we list all models with the representation assignments given in Table~\ref{cont} and higher $k_{Y{\rm max}}$ but not larger than 10 in Table~\ref{mW}. As one can see, only the textures ${\bf B_2}$, ${\bf B_4}$ and ${\bf C}$ can be achieved with the exact representation assignments given in Table~\ref{cont}. Nevertheless, the textures ${\bf A_{1,2}}$ and ${\bf B_{1,3}}$ can be obtained by exchanging or alternating the representation assignments and weights of the chiral superfields in these models leading to ${\bf B_{2,4}}$ and ${\bf C}$. For example, all the models generating the texture ${\bf A_2}$ can be obtained by exchanging the representation assignments and weights of $L_1$ and $L_3$ together with those of $E^c_1$ and $E^c_3$ in all these models giving the texture ${\bf B_4}$, which is denoted as
\begin{eqnarray}
	L_1 \leftrightarrow L_3 \;,\; E^c_1 \leftrightarrow E^c_3 \quad \Rightarrow \quad {\bf B_4} \rightarrow {\bf A_2} \;,
	\label{b1}
\end{eqnarray} 
where we use $\phi_i \leftrightarrow \phi_j$ (or $\phi_i \rightarrow \phi_j \rightarrow \phi_k \rightarrow \phi_i$) to represent exchanging the representation assignments and weights of $\phi_i$ and $\phi_j$ (or alternating those of $\phi_i$, $\phi_j$ and $\phi_k$), with $\phi$ being the chiral superfield. Then similarly, we have
\begin{eqnarray}
&L_1& \leftrightarrow L_2 \;,\; E^c_1 \leftrightarrow E^c_2 \quad \Rightarrow \quad {\bf B_2} \rightarrow {\bf B_2} \;,
\nonumber
\\
&L_1& \leftrightarrow L_3 \;,\; E^c_1 \leftrightarrow E^c_3 \quad \Rightarrow \quad {\bf B_4} \rightarrow {\bf A_2} \;,
\nonumber
\\
&L_2& \leftrightarrow L_3 \;,\; E^c_2 \leftrightarrow E^c_3 \quad \Rightarrow \quad {\bf B_2} \rightarrow {\bf B_1} \;,\; {\bf B_4} \rightarrow {\bf B_3} \;,\; {\bf C} \rightarrow {\bf C} \;,
\nonumber
\\
&L_1& \rightarrow L_2 \rightarrow L_3 \rightarrow L_1 \;,\; E^c_1 \rightarrow E^c_2 \rightarrow E^c_3 \rightarrow E^c_1 \quad \Rightarrow \quad {\bf B_4} \rightarrow {\bf A_1} \;,
\nonumber
\\
&L_1& \rightarrow L_3 \rightarrow L_2 \rightarrow L_1 \;,\; E^c_1 \rightarrow E^c_3 \rightarrow E^c_2 \rightarrow E^c_1 \quad \Rightarrow \quad {\bf B_2} \rightarrow {\bf B_1} \;,
\label{b2}
\end{eqnarray}
which give all the other possible models.

\begin{table}[htp]
	\centering
	\vspace{-0.2cm}
	\caption{The representation assignments and modular weights for the chiral supermultiplets.}
	\setlength{\tabcolsep}{0.3cm}
	\vspace{0.2cm}
	\begin{tabular}{|c|c|c|c|c|c|c|c|c|c|c||c|}
		\hline
		\multicolumn{3}{|c|}{} & $L_1$ & $L_2$ & $L_3$ & $E^c_1$ & $E^c_2$ & $E^c_3$ & $H_u$ & $H_d$ & $k_{Y{\rm max}}$ \\
		\hline
		\multicolumn{3}{|c|}{$A_4$} & ${\bf 1}$ &  ${\bf 1^\prime}$ &  ${\bf 1^{\prime\prime}}$ &  ${\bf 1}$ &  ${\bf 1^{\prime\prime}}$ &${\bf 1^\prime}$ &  ${\bf 1}$ &  ${\bf 1}$ & --- \\
		\hline
		\multirow{10}{*}{$-k$}& \multirow{5}{*}{${\bf B_2}$}& \ding{172} & $-2$ & $-2$ & $-2$ & $-4$, $2$ & $-4$, $2$ & $-4$, $2$ & 0 & 0 & 4, 6 \\
		\cline{3-12}
		& & \ding{173} & $-2$ & $-4$ & $-2$ & 2 & $-4$, $0$, $4$ & $-2$, $2$ & 0 & 0 & 8\\
		\cline{3-12}
		& & \ding{174} & $-5$ & $-5$ & $-3$ & 1, 5 & $-1$, 5 & $-1$, 3 & 0 & 0 & 10\\
		\cline{3-12}
		& & \ding{175} & $-5$ & $-5$ & $-5$ & $-1$, 5 & $-1$, 5 & $-1$, 5 & 0 & 0 & 10\\
		\cline{3-12}
		& & \ding{176} & $-5$ & $-5$ & 1 & $-1$, 1, 5 & $-1$, 5 & $-1$ & 0 & 0 & 10\\
		\cline{2-12}
		& \multirow{3}{*}{${\bf B_4}$}& \ding{172} & $-4$ & $-4$ & $-2$ & 0, 4 & $-2$, 4 & $-2$, 2 & 0 & 0 & 8 \\
		\cline{3-12}
		& & \ding{173} & $-3$ & $-5$ & $-3$ & 3 & $-3$, 1, 5 & $-1$, 3 & 0 & 0 & 10\\
		\cline{3-12}
		& & \ding{174} & $-3$ & $-5$ & 1 & $-1$, 3 & $-3$, 1, 5 & $-1$ & 0 & 0 & 10\\
		\cline{2-12}
		& \multirow{2}{*}{${\bf C}$}& \ding{172} & $-5$ & $-3$ & $-3$ & $-3$, 1, 5 & $-1$, 3 & $3$ & 0 & 0 & 10 \\
		\cline{3-12}
		& & \ding{173} & $-5$ & $-3$ & $-5$ & $-1$, 5 & $-1$, 3 & 1, 5 & 0 & 0 & 10\\
		\hline
	\end{tabular}
	\label{mW}
\end{table}
\vspace{-0.5cm}

\subsection{The Type-\uppercase\expandafter{\romannumeral1} Seesaw Mechanism}
In the case where the type-\uppercase\expandafter{\romannumeral1} seesaw mechanism is used to generate neutrino masses, the situation is much more complicated. Given the right-handed neutrino mass matrix in Eq.~(\ref{rm}), all the Dirac neutrino mass matrices which can give two-zero textures of $M_\nu$ are listed in Table~\ref{md} before considering any symmetries. Only four-zero and five-zero textures of $M_{\rm D}$ can be realized with the modular $A_4$ symmetry. Since there are so many models leading to two-zero textures of $M_\nu$, we do not list the representation assignments and weights for every model, as what we have done in the case where neutrino masses come from the Weinberg operation. Instead, for a given two-zero texture of $M_\nu$ and a specific representation assignment, we only list the number of different models (in which modular weights for the chiral superfields are different) and the achieved textures of $M_{\rm D}$ together with the corresponding minimal $k^{\prime}_{Y{\rm max}}$. 
\begin{table}[h]
	\centering
	\vspace{-0.2cm}
	\caption{All the possible structures of the Dirac neutrino mass matrix $M_{\rm D}$ giving two-zero textures of the Majorana neutrino mass matrix $M_\nu$ with right-handed neutrino mass matrix $M_{\rm R}$ given in Eq.~(\ref{rm}) before considering any symmetries.}
	\vspace{0.2cm}
	\resizebox{\textwidth}{20mm}{
		\begin{tabular}{|c|c|c|c|c|c|c|c|c|c|c|}
			\hline
			No. & 1 & 2 & 3 & 4 & 5 & 6 & 7 & 8 & 9 & 10 \\
			\hline
			${\bf A_1}$ & 
			$\left(\begin{smallmatrix} 0 & 0 & \times \\ \times & 0 & \times \\ \times & \times & \times \end{smallmatrix}\right)$ & 
			$\left(\begin{smallmatrix} 0 & \times & 0 \\ \times & \times & 0 \\ \times & \times & \times \end{smallmatrix}\right)$ & 
			$\left(\begin{smallmatrix} 0 & 0 & \times \\ \times & 0 & 0 \\ \times & \times & \times \end{smallmatrix}\right)$ & 
			$\left(\begin{smallmatrix} 0 & 0 & \times \\ \times & 0 & \times \\ 0 & \times & \times \end{smallmatrix}\right)$ & 
			$\left(\begin{smallmatrix} 0 & 0 & \times \\ \times & 0 & \times \\ \times & \times & 0 \end{smallmatrix}\right)$ & 
			$\left(\begin{smallmatrix} 0 & \times & 0 \\ \times & 0 & 0 \\ \times & \times & \times \end{smallmatrix}\right)$ & 
			$\left(\begin{smallmatrix} 0 & \times & 0 \\ \times & \times & 0 \\ 0& \times & \times \end{smallmatrix}\right)$ & 
			$\left(\begin{smallmatrix} 0 & \times & 0 \\ \times & \times & 0 \\ \times & 0 & \times \end{smallmatrix}\right)$& 
			$\left(\begin{smallmatrix} 0 & 0 & \times \\ \times & 0 & 0 \\ \times & \times & 0 \end{smallmatrix}\right)$ & 
			$\left(\begin{smallmatrix} 0 & \times & 0 \\ \times & 0 & 0 \\ \times & 0 & \times \end{smallmatrix}\right)$    \\
			\hline
			${\bf A_2}$ & 
			$\left(\begin{smallmatrix} 0 & 0 & \times \\ \times & \times & \times \\ \times & 0 & \times  \end{smallmatrix}\right)$ & 
			$\left(\begin{smallmatrix} 0 & \times & 0 \\ \times & \times & \times \\ \times & \times & 0 \end{smallmatrix}\right)$ & 
			$\left(\begin{smallmatrix} 0 & 0 & \times \\ \times & \times & \times \\ \times & 0 & 0 \end{smallmatrix}\right)$ & 
			$\left(\begin{smallmatrix} 0 & 0 & \times \\ 0 & \times & \times \\ \times & 0 & \times \end{smallmatrix}\right)$ & 
			$\left(\begin{smallmatrix} 0 & 0 & \times \\ \times & \times & 0 \\ \times & 0 & \times \end{smallmatrix}\right)$ & 
			$\left(\begin{smallmatrix} 0 & \times & 0 \\ \times & \times & \times \\ \times & 0 & 0 \end{smallmatrix}\right)$ & 
			$\left(\begin{smallmatrix} 0 & \times & 0 \\ 0& \times & \times \\ \times & \times & 0 \end{smallmatrix}\right)$ & 
			$\left(\begin{smallmatrix} 0 & \times & 0 \\ \times & 0 & \times \\ \times & \times & 0 \end{smallmatrix}\right)$& 
			$\left(\begin{smallmatrix} 0 & 0 & \times \\ \times & \times & 0 \\ \times & 0 & 0 \end{smallmatrix}\right)$ & 
			$\left(\begin{smallmatrix} 0 & \times & 0 \\ \times & 0 & \times \\ \times & 0 & 0 \end{smallmatrix}\right)$    \\
			\hline
			${\bf B_3}$ & 
			$\left(\begin{smallmatrix} \times & 0 & \times \\ 0 & 0 & \times \\ \times & \times & \times \end{smallmatrix}\right)$ & 
			$\left(\begin{smallmatrix} \times & \times & 0 \\ 0 & \times & 0 \\ \times & \times & \times \end{smallmatrix}\right)$ & 
			$\left(\begin{smallmatrix} \times & 0 & 0 \\ 0 & 0 & \times \\ \times & \times & \times \end{smallmatrix}\right)$ & 
			$\left(\begin{smallmatrix} \times & 0 & \times \\ 0 & 0 & \times \\ 0 & \times & \times \end{smallmatrix}\right)$ & 
			$\left(\begin{smallmatrix} \times & 0 & \times \\ 0 & 0 & \times \\ \times & \times & 0 \end{smallmatrix}\right)$ & 
			$\left(\begin{smallmatrix} \times & 0 & 0 \\ 0 & \times & 0 \\ \times & \times & \times \end{smallmatrix}\right)$ & 
			$\left(\begin{smallmatrix} \times & \times & 0 \\ 0 & \times & 0 \\ 0 & \times & \times \end{smallmatrix}\right)$ & 
			$\left(\begin{smallmatrix} \times & \times & 0 \\ 0 & \times & 0 \\ \times & 0 & \times \end{smallmatrix}\right)$& 
			$\left(\begin{smallmatrix} \times & 0 & 0 \\ 0 & 0 & \times \\ \times & \times & 0 \end{smallmatrix}\right)$ & 
			$\left(\begin{smallmatrix} \times & 0 & 0 \\ 0 & \times & 0 \\ \times & 0 & \times \end{smallmatrix}\right)$    \\
			\hline
			${\bf B_4}$ & 
			$\left(\begin{smallmatrix} \times & 0 & \times \\ \times & \times & \times \\ 0 & 0 & \times \end{smallmatrix}\right)$ & 
			$\left(\begin{smallmatrix} \times & \times & 0 \\ \times & \times & \times \\ 0 & \times & 0 \end{smallmatrix}\right)$ & 
			$\left(\begin{smallmatrix} \times & 0 & 0 \\ \times & \times & \times \\ 0 & 0 & \times \end{smallmatrix}\right)$ & 
			$\left(\begin{smallmatrix} \times & 0 & \times \\ 0 & \times & \times \\ 0 & 0 & \times \end{smallmatrix}\right)$ & 
			$\left(\begin{smallmatrix} \times & 0 & \times \\ \times & \times & 0 \\ 0 & 0 & \times \end{smallmatrix}\right)$ & 
			$\left(\begin{smallmatrix} \times & 0 & 0 \\ \times & \times & \times \\ 0 & \times & 0 \end{smallmatrix}\right)$ & 
			$\left(\begin{smallmatrix} \times & \times & 0 \\ 0 & \times & \times \\ 0 & \times & 0 \end{smallmatrix}\right)$ & 
			$\left(\begin{smallmatrix} \times & \times & 0 \\ \times & 0 & \times \\ 0 & \times & 0 \end{smallmatrix}\right)$& 
			$\left(\begin{smallmatrix} \times & 0 & 0 \\ \times & \times & 0 \\ 0 & 0 & \times \end{smallmatrix}\right)$ & 
			$\left(\begin{smallmatrix} \times & 0 & 0 \\ \times & 0 & \times \\ 0 & \times & 0 \end{smallmatrix}\right)$    \\
			\hline
			${\bf C}$ & 
			$\left(\begin{smallmatrix} \times & \times & \times \\ 0 & 0 & \times \\ 0 & \times & 0 \end{smallmatrix}\right)$ & 
			$\left(\begin{smallmatrix} \times & \times & \times \\ 0 & \times & 0 \\ 0 & 0 & \times \end{smallmatrix}\right)$ & 
			$\left(\begin{smallmatrix} 0 & \times & \times \\ 0 & 0 & \times \\ 0 & \times & 0 \end{smallmatrix}\right)$ & 
			$\left(\begin{smallmatrix} 0 & \times & \times \\ 0 & \times & 0 \\ 0 & 0 & \times \end{smallmatrix}\right)$ & --- & --- & --- & --- & --- & --- \\
			\hline
	\end{tabular}}
	\label{md}
\end{table}
\begin{table}[h!]
	\centering
	\vspace{-0.2cm}
	\caption{All possible models leading to two-zero textures of $M_\nu$ with maximal weight not larger than 10. The ``{\bf 1}" stands for the representation assignments given in Table~\ref{cont} exactly and ``$ i \leftrightarrow j $" (or ``$ i \rightarrow j \rightarrow k \rightarrow i$") represents exchanging the representation assignments of $L_i$ and $L_j$ as well as those of $E^c_i$ and $E^c_j$ (or alternating representation assignments of $L_{i,j,k}$ and $E^{c}_{i,j,k}$ in the corresponding order) in Table~\ref{cont}. The entries $x,y,z$ in a triad $(x,y,z)$ stand for the identifier of achieved $M_{\rm D}$ given in Table~\ref{md}, the number of different models and the minimal $k^{\prime}_{Y{\rm max}}$ among those models, respectively. }
	\vspace{0.2cm}
	\begin{tabular}{|c|c|c|c|c|c|c|}
		\hline
		& {\bf 1} & $1 \leftrightarrow 2$ & $1 \leftrightarrow 3$ & $2 \leftrightarrow 3$ & $1 \rightarrow 2 \rightarrow 3 \rightarrow 1$ & $1 \rightarrow 3 \rightarrow 2 \rightarrow 1$ \\
		\hline
		${\bf A_1}$ & (5, 6, 10) & \tabincell{c}{ (3, 28, 8) \\(4, 14, 8) \\(5, 56, 6) \\(9, 78, 6) } & (8, 6, 10)  & --- &  \tabincell{c}{ (6, 12, 8) \\(8, 6, 10) \\ (10, 100, 8)} & \tabincell{c}{ (3, 4, 10) \\(8, 30, 8) \\(9, 4, 10) \\(10, 6, 8)}  \\
		\hline
		${\bf A_2}$ & --- & \tabincell{c}{(3, 4, 10) \\(8, 30, 8) \\(9, 4, 10) \\(10, 6, 8)} & \tabincell{c}{(6, 12, 8) \\(8, 6, 10) \\(10, 100, 8)} & (5, 6, 10) & (8, 6, 10) & \tabincell{c}{(3, 28, 8) \\(4, 14, 8) \\(5, 56, 6)  \\(9, 78, 6)}  \\
		\hline
		${\bf B_3}$ & \tabincell{c}{(3, 28, 8)\\ (4, 14, 8) \\(5, 56, 6)  \\(9, 78, 6)} & (5, 6, 10) & \tabincell{c}{(3, 4, 10) \\(8, 30, 8) \\(9, 4, 10) \\(10, 6, 8)} & \tabincell{c}{(6, 12, 8) \\(8, 6, 10) \\(10, 100, 8)} & --- & (8, 6, 10) \\
		\hline 
		${\bf B_4}$ & \tabincell{c}{ (6, 12, 8) \\(8, 6, 10) \\ (10, 100, 8)} & (8, 6, 10) & ---  & \tabincell{c}{ (3, 28, 8) \\(4, 14, 8) \\(5, 56, 6) \\(9, 78, 6) } & \tabincell{c}{ (3, 4, 10) \\(8, 30, 8) \\(9, 4, 10) \\(10, 6, 8)} & (5, 6, 10)  \\
		\hline
		${\bf C}$  & \tabincell{c}{(1, 16, 8) \\(3, 192, 8)} & (3, 862, 6) & \tabincell{c}{(3, 18, 10) \\(4, 18, 10)} & \tabincell{c}{(2, 16, 8) \\(4, 192, 8)} & \tabincell{c}{(3, 18, 10) \\(4, 18, 10)} & (4, 862, 6) \\
		\hline
	\end{tabular}
	\label{mS}
\end{table}

All the possible models are listed in Table~\ref{mS}. The first row shows the representation assignments for the chiral superfields, where ``{\bf 1}" stands for the representation assignments given in Table~\ref{cont} exactly and ``$ i \leftrightarrow j $" (or ``$ i \rightarrow j \rightarrow k \rightarrow i$") represents exchanging the representation assignments of $L_i$ and $L_j$ as well as those of $E^c_i$ and $E^c_j$ (or alternating representation assignments of $L_{i,j,k}$ and $E^{c}_{i,j,k}$ in the corresponding order) in Table~\ref{cont}. And the entries $x,y,z$ in a triad $(x,y,z)$ stand for the identifier of the achieved $M_{\rm D}$ given in Table~\ref{md}, the number of different models and the minimal $k^{\prime}_{Y{\rm max}}$ among those models, respectively.  For example, the triad $(5, 6, 10)$ in the fifth column of the third row means that to achieve the texture ${\bf A_2}$ with the representation assignments given by exchanging those of $L_2$ and $L_3$ together with those of $E^c_2$ and $E^c_3$ in Table~\ref{cont}, there are totally six different models with different weight assignments and the minimal $k^{\prime}_{Y{\rm max}} = 10$, and all such models give the texture of $M_{\rm D}$ shown in the sixth column and the third row of Table~\ref{md}. One can see that there are similar relations among the models leading to different two-zero textures of $M_\nu$ like those in Eq.~(\ref{b2}). These are induced by the permutation relations of different two-zero textures of $M_\nu$ and textures of the corresponding $M_{\rm D}$.

\end{appendices}

\end{document}